\documentclass[sigconf, screen]{acmart}
\AtBeginDocument{%
  }

\setcopyright{acmlicensed}
\copyrightyear{2024}
\acmYear{2024}
\acmDOI{XXXXXXX.XXXXXXX}
\usepackage{textcomp}
\usepackage{xspace}
\usepackage{url}
\usepackage{hyperref}
\usepackage{array}
\usepackage{enumitem}
\begin{document}

\title{Integrating DAST in Kanban and CI/CD: A Real-World Security Case Study}
\author{Arpit Thool}
\email{arpitthool@vt.edu}
\orcid{0009-0003-8132-2887}
\affiliation{%
  \institution{Virginia Tech}
  \city{Blacksburg}
  \state{Virginia}
  \country{USA}
}

\author{Chris Brown}
\email{dcbrown@vt.edu}
\orcid{0000-0002-6036-4733}
\affiliation{%
  \institution{Virginia Tech}
  \city{Blacksburg}
  \state{Virginia}
  \country{USA}
}









\begin{abstract}
Modern development methodologies, such as Kanban and continuous integration and continuous deployment (CI/CD), are critical for web application development---as software products must adapt to changing requirements and deploy products to users quickly. As web application attacks and exploited vulnerabilities are rising, it is increasingly crucial to integrate security into modern development practices. Yet, the iterative and incremental nature of these processes can clash with the sequential nature of security engineering. Thus, it is challenging to adopt security practices and activities in modern development practices. Dynamic Application Security Testing (DAST) is a security practice within software development frameworks that bolsters system security. This study delves into the intersection of Agile development and DAST, exploring how a software organization attempted to integrate DAST into their Kanban workflows and CI/CD pipelines to identify and mitigate security vulnerabilities within the development process. Through an action research case study incorporating interviews among team members, this research elucidates the challenges, mitigation techniques, and best practices associated with incorporating DAST into Agile methodologies from developers' perspectives. We provide insights into integrating security practices with modern development, ensuring both speed and security in software delivery.

\end{abstract}

\ccsdesc[500]{Software and its engineering~Agile software development}
\ccsdesc[500]{Security and privacy~Software security engineering}
\ccsdesc[300]{Security and privacy~Web application security}
\ccsdesc[500]{Security and privacy~Social aspects of security and privacy}

\keywords{Agile, Case Study, Dynamic Application Security Testing (DAST)}


\maketitle

\section{Introduction}

Organizations are increasingly deploying \textit{web applications}, software accessed through web browsers~\cite{jazayeri2007some}, to deliver online services to users across domains~\cite{andreessen2011software}. To implement these products, development teams often rely on a variety of modern development practices. For example, Agile software development methodologies are the most popular development approach~\cite{hoda2018rise}---reportedly adopted by over 70\% of companies in the United States~\cite{techreport}. Studies show web applications are often developed using iterative Agile processes~\cite{hu2008agile}, such as Kanban~\cite{kumar2021implementing}. Further, continuous integration and continuous delivery (CI/CD) processes~\cite{cidd} are often used to build, test, and deploy code changes to users \cite{rahman2015synthesizing}. Studies show CI/CD practices are commonly adopted to support web application development~\cite{naveen2023efficient}.

Yet, the majority of security breaches can be exploited through web applications~\cite{verizon,nsa}---making web application security increasingly critical as online threats evolve and expand~\cite{alzahrani2017web}. Thus, integrating security into modern development methodologies is paramount, particularly when software deals with sensitive information~\cite{chowdhury2016recent} and data protection requirements~\cite{boltz2022model}. For instance, web applications are increasingly collecting data from users~\cite{bucklin2009click}. However, modern software development practices are difficult to secure. For instance, prior work suggests traditional Security Engineering is sequential in nature~\cite{rindell2018aligning}, conflicting with Agile workflows. Research also shows security activities are challenging to integrate in CI/CD pipelines~\cite{rajapakse2021empirical,bajpai2022secure}. This creates a critical gap: while modern software must adapt quickly, it must also be secure and robust.  


Dynamic Application Security Testing (DAST) is a security practice that automates security testing in web applications. DAST tools are widely used to assess the security of web applications~\cite{dencheva2022comparative} by dynamically simulating attacks and evaluating applications externally, mimicking the actions of a malicious attacker~\cite{sharma2021review}. 
DAST tools generate security reports that rank vulnerabilities based on severity, enabling development teams to address critical issues before deployment~\cite{aydos2022security}. 

Traditional security practices, with their reliance on lengthy planning and heavy documentation~\cite{riisom2018software}, can be challenging to implement in modern fast-moving, iterative environments. However, developers' perceptions of the real-world integration of DAST into Kanban and CI/CD processes remains an under-explored area---particularly in industry contexts where speed and security are often at odds. Therefore, this study explores practitioner perspectives on integrating DAST into a Kanban team. We investigate the perceived impacts and challenges of integrating automated security testing to detect and mitigate vulnerabilities during development and maintenance. To accomplish this, we aim to answer the following research questions (RQs):

\begin{itemize}[topsep=0pt]
    \item[]\textbf{RQ1} \label{RQ1-willing} How willing are Kanban team members to adopt and continue using DAST in their development process, and what factors influence their willingness?

    
    \item[]\textbf{RQ2} \label{RQ2-impact} What are the key challenges and perceived impacts of integrating DAST into Kanban workflows?
    
    
    \item[]\textbf{RQ3} \label{RQ3-improvements} What improvements can be made to better integrate security practices into Kanban?
    
    \item[]\textbf{RQ4} \label{RQ4-compare} How does the use of DAST compare to other security practices familiar to team members?
    
\end{itemize}

We seek to answer these RQs through an action research case study~\cite{petersen2014action}, involving a real-world software development team that undertook the initiative to adopt DAST in modern methodologies---specifically Kanban and CI/CD processes. The first author was embedded as a member of this development team and led the efforts to implement DAST scans as part of their CI/CD pipeline. Our study focuses on aspects of DAST integration in Kanban and CI/CD processes, seeking to understand the perspectives and experiences of different stakeholders within the team---developers, testers, data analysts, and managers---throughout the process. To gain a deeper understanding, qualitative interviews were conducted with ten team members, providing valuable insights into the challenges and impact of integrating security practices in modern workflows. 

Our findings make three key contributions: (1) we present a real-world case study on how DAST was integrated into a Kanban-based web development team, offering insights into the technical and organizational challenges involved; (2) we provide empirical evidence from qualitative interviews with practitioners across various roles, highlighting the factors influencing the willingness to adopt and continue using DAST in Kanban and Agile environments; (3) we offer practical recommendations and lessons learned for software development teams in industry to improve the integration of DAST into development workflows. By addressing these points, our work contributes to a broader understanding of how to effectively incorporate security into modern development processes.

\section{Background}
\label{section-background}

\subsection{Kanban}
\label{section-background-kanban}

Kanban is a widely adopted Agile framework~\cite{winska2020software} that emphasizes visualizing workflows by using cards\cite{junior2010variations}, limiting work in progress (WIP)~\cite{al2015kanban}, and promoting continuous delivery without the rigid time-boxed iterations commonly associated with other Agile methodologies, such as Scrum~\cite{hofmann2018development, cvetkovic2017enhancing}. Originating from lean manufacturing principles\cite{hosseini2015lean}, Kanban emphasizes optimizing efficiency by encouraging teams to pull tasks into their workflows based on capacity~\cite{yacoub2016new} rather than adhering to a predefined sprint schedule. This adaptability makes Kanban particularly attractive for teams that require flexibility in managing varying workloads and priorities~\cite{malakar2021agile}. The Kanban board is typically divided into columns representing different stages of the development process (e.g., ``To Do,'' ``In Progress,'' ``In Review,'' ``Done'')~\cite{saltz2020exploring}. By efficiently moving cards/tasks through the pipeline, teams can continuously release features and updates, fostering a more dynamic development environment. The emphasis on WIP limits ensures that teams do not overload themselves~\cite{cvetkovic2017enhancing}, thereby reducing bottlenecks and improving workflow.

\label{section-background-DAST}
\subsection{Dynamic Application Security Testing (DAST)}
Dynamic Application Security Testing (DAST) is a technique used to evaluate the security of web applications and APIs in a live environment~\cite{pan2019interactive}. Unlike Static Application Security Testing (SAST), which focuses on analyzing the application's source code~\cite{dupont2021incremental}, DAST simulates an attacker by interacting with the application from the outside. This real-time approach enables DAST tools to uncover runtime vulnerabilities, including issues like SQL injection, cross-site scripting (XSS), and authentication flaws~\cite{sakaoglu2023kartal}. Various commercial and open-source DAST tools are available, enabling organizations to identify and address security weaknesses in web applications and APIs, thereby strengthening their overall security posture.
DAST operates as a \textbf{Black-box Testing} testing method~\cite{mateo2020combining}, meaning it does not require access to source code, interacting with the application through external interfaces like web forms and APIs~\cite{sakaoglu2023kartal}. 
DAST performs \textbf{Dynamic Analysis} by sending requests and inputs to the application while it is running, uncovering vulnerabilities that may only emerge under specific conditions~\cite{mateo2020combining}.
DAST tools \textbf{Simulate Real-World} attacks~\cite{casanova2021application} by sending malicious payloads to probe the application for security flaws, helping organizations evaluate how well their applications would withstand actual cyber threats. Once a scan is complete, DAST tools generate a report categorizing vulnerabilities by severity.

\section{Case Study Setting}\label{sec:team}

Our case study was conducted within the Identity Service team of an organization under their broader Information Technology (IT) division. The organization's name has been anonymized to protect the integrity of ongoing discussions about security protocols. The Identity Services team plays a critical role in ensuring the secure management of digital identities by providing authentication and authorization for the organization. These services facilitate safe access to resources, applications, and data, supporting vital functions such as operations, research, and outreach. The Identity Services team is committed to maintaining and continuously improving identity and access services, focusing on high-security standards, user-friendly design, and adaptability to industry changes. 

\subsection{Team Organization}

The team consists of 23 members and is formed by three distinct groups\label{Team-three-groups}: Developers \& DevOps, Testers, and Identity and Access Management (IAM) Analysts, all of which collaborate to provide agile, secure, and seamless digital identity solutions across the organization’s ecosystem. The team manages key applications, which include a web Single-Sign-On (SSO) service, a web application for directory administration, an account creation system, and an enterprise directory for identity management. Each group plays a crucial role in the software development process, with Developers \& DevOps focusing on code development and infrastructure, Testers ensuring the quality of the product, and Analysts providing insights and feedback based on system requirements and performance metrics. All team members worked remotely and in the same timezone.

\subsection{Development Process} The team follows the Kanban framework, emphasizing continuous flow and limiting work in progress to optimize efficiency. All team members are familiar with Agile and Kanban. For the team's specific implementation, stand-up calls are held every morning, serving as a touchpoint for members to discuss progress, blockers, and plans for the day. After the stand-up, additional discussions are encouraged if team members have specific topics or issues they wish to address. The work is managed through three Kanban boards, one for each group. 
Combining daily stand-up calls followed by after-Kanban discussion
practices balances flexibility and accountability. It is a suitable environment for integrating additional security practices~\cite{bartsch2011practitioners}, such as DAST, into the Agile SDLC. The team uses GitLab for version control and to run and configure their CI/CD pipeline.


\section{Methodology}

We conducted an action research case study~\cite{petersen2014action,staron2020action} with a real-world software development team to gain detailed insights from developers on the challenges and impacts of integrating DAST tools in Agile development. Researchers suggest case studies are valuable in software engineering research to observe phenomena in relevant industry contexts~\cite{runeson2009guidelines,WOHLIN2021106514}. Further, previous studies have similarly leveraged case studies to observe various software engineering phenomena in practice (i.e.,~\cite{lwakatare2021experiences, davila2024industry, voggenreiter2024automated}). Our study incorporated the five iterative steps of the action research cycle~\cite{petersen2014action,staron2020action}: problem diagnosis; action planning, action taking; evaluation; and specifying learning. We describe these steps in detail below.

\subsection{Problem Diagnosis}

Problem diagnosis focuses on identifying and describing the problem in industrial contexts~\cite{petersen2014action}. Our prior work suggests software development teams utilizing Agile development processes often face challenges adopting security activities into their workflow~\cite{arpit}. In addition, studies show UIs---in particular web applications---are challenging to develop~\cite{myers1995state}, test~\cite{memon2002UI}, and secure~\cite{mitropoulos2017defending}. For instance, many security vulnerabilities, such as cross-site scripting, access management issues, and information leakage, can be exploited through UIs~\cite{luo2017hindsight}. To mitigate these exploits, DAST tools have been introduced to assess web application security by simulating attacks to its user interface~\cite{dencheva2022comparative}. However, there is limited knowledge exploring the adoption of DAST in industry contexts. 

The primary goal of the case study team was to explore possible methods of incorporating security testing into the team's CI/CD pipeline for a selected web application based in Java. The team lacked sufficient security testing, primarily relying on manual and ad hoc efforts to verify the robustness of their system---leading them to explore DAST integration. This task, created in the project backlog and later assigned to an individual engineer, marked the beginning of the team's DAST integration effort. The task was eventually moved to the ``Implement'' column on the Kanban board, where it became the focus of this project. Integrating DAST into the team's workflow began as a research initiative within the testing group, of which the first author is a member. This research initiative emerged from the testing group's interest in adding security tests to their existing suite of test sets. They recognized the limitations of their current test suite, which had no security tests, and sought to investigate how automated security testing could be integrated into their development workflows.

The scope of our case study investigates how a real-world software development team (see Section~\ref{sec:team}) adopts DAST in Kanban and CI/CD development processes to secure web applications. Our study employed focused observation~\cite{stausberg2011structured}, a field technique where an observer concentrates on specific phenomena in a natural setting to provide in-depth insights on DAST integration in Agile practices. We implemented this approach with a high degree of interaction by the researchers and low awareness of being observed by the team~\cite{runeson2009guidelines}. The first author was actively a member of the team during the integration of DAST within their workflows to secure systems, collaborating with the testing team and all 23 members of the team, and a manager to support DAST integration in their project workflow. The first author spent three months with the team, taking part in normal work activities, including stand-up meetings and after-Kanban discussions. Each activity was observed in real-time in-person, through video conferencing tools (i.e., Zoom and Microsoft Teams), and in communications platforms (i.e., Slack). Observations were scheduled during regular team activities to avoid disruptions and allow the team to continue work naturally. Participant confidentiality and anonymity were assured, and this research initiative was approved by the team manager.  

\subsection{First Iteration}

\subsubsection{Action Planning}

This phase explores various methods for solving the identified problem~\cite{petersen2014action}. The plan was co-created with the researcher and the team based on the needs and constraints of the case study organization---i.e., its infrastructure and resources. The team desired the following use case to automate DAST scans in their workflow: \textit{(a)} the system must have the capability to act as an authorized user through an API; \textit{(b)} Once logged in, control must be transferred to the DAST tool; and \textit{(c)} The system, acting as an authenticated user, must conduct the necessary security scans to detect for vulnerabilities and ensure compliance. To fulfill this requirement, it was essential to identify a DAST solution that could be programmatically controlled through code, enabling seamless integration into the CI/CD pipeline. Based on these requirements, we explored eight potential DAST solutions---identifying OWASP ZAP\footnote{\url{https://www.zaproxy.org/}} as the most suitable tool. Zed Attack Proxy (ZAP) is a free, open-source, cross-platform web application security scanner maintained by the Open Web Application Security Project (OWASP)~\cite{sinchana2019performance}. It provides automated scanners for vulnerability assessment and an interactive manual testing environment for analysis. ZAP satisfied the requirement of being callable via code through its local API, making it an ideal choice for integrating into the team's existing workflow.

\subsubsection{Action Taking}

This phase involves implementing the planned actions~\cite{petersen2014action}---in this case, integrating the selected DAST tool ZAP to secure the team's product within their Agile development practices. The researcher introduced the action as a \textit{direct intervention}~\cite{staron2020action}, directly changing the organization's operations. The engineer responsible for this task implemented a Java program that leveraged the in-house Java software to sign in to the web application with authorized user credentials. Once signed in, the program extracted the necessary authentication token and cookies from the session and transferred them to the ZAP application environment. This allowed ZAP to access the web application authentically, ensuring the security scans were conducted with the necessary privileges.
The Java program then communicated with the locally installed ZAP instance through its API, specifying the scans to be performed. Once the scans were completed, the program generated a security report, providing the team with detailed insights into any vulnerabilities identified within the web application.
This implementation allowed the testing team to integrate automated security scans into the CI/CD pipeline without disrupting the existing workflow, ensuring that security remained a core focus throughout the software development lifecycle.

After successfully developing the initial Java program to conduct DAST scans using ZAP, the team tested it on several web applications. While the program worked as expected for simpler applications built with HTML and CSS, we encountered issues when testing JavaScript-heavy web applications:

\textbf{Dynamic Content:} ZAP struggled to handle dynamic content generated by modern JavaScript frameworks, which limited its effectiveness for more complex applications.

\textbf{HTTP Protocol:} Additionally, at the time of implementation, ZAP only supported the HTTP/2 protocol, while many of our web applications had already migrated to the newer HTTP/3 standard. This limitation further constrained the utility of ZAP in our testing environment, prompting us to explore alternative DAST solutions.

\subsection{Second Iteration}

\subsubsection{Action Planning}

After evaluating several options, including WebInspect and BurpSuite, we chose to move forward with BurpSuite\footnote{\url{https://portswigger.net/burp/pro}} based on its feature set and the input from the rest of the team. Burp Suite, developed by 
PortSwigger, is a commercial web vulnerability scanner. Similar to ZAP, it offers both automated and manual testing capabilities. Burp Suite features a variety of inbuilt extensions, such as Proxy, Intruder, and Repeater, to facilitate penetration testing~\cite{kore2022burp}.

\subsubsection{Action Taking}

The team initially experimented with the free version of BurpSuite to assess its capability for running authenticated scans. With assistance from BurpSuite’s support team, we obtained the necessary scripts and configurations to perform security scans as an authenticated user, allowing us to achieve the desired functionality.
The transition to BurpSuite provided a more robust solution for testing our JavaScript-heavy applications and resolved the protocol compatibility issues we had encountered with ZAP. This decision marked an important step in ensuring the DAST implementation was effective and aligned with the team's web application requirements. After determining that BurpSuite met the requirements, we obtained a subscription to BurpSuite Pro with additional features and support. However, in subsequent after-Kanban meetings following DAST integration, several issues arose:

\textbf{Timing:} During an after-Kanban discussion, the team agreed that DAST scans would be conducted quarterly to balance security concerns with development timelines. However, this conflicts with CI/CD and Agile values, which suggest automated test suites should be run frequently to ensure and maintain software quality~\cite{rahman2015synthesizing}.

\textbf{Lack of Bandwidth:} Team members lacked the bandwidth to address all security alerts generated by DAST. After further discussion, it was agreed that only medium and high-severity alerts would be prioritized for resolution. This decision ensured critical security vulnerabilities were addressed without overwhelming the development team.


\subsection{Evaluation}

The evaluation phase seeks to understand the effects of the captured actions~\cite{petersen2014action}. We conducted \textit{qualitative interviews} with developers to reflect on their experiences integrating DAST into their project Agile workflow. Our interview protocol was approved by the Institutional Review Board (IRB).

\subsubsection{Data Collection}

\begin{table}[t] \label{table-participants}
    \centering
    \caption{Interview Participants}

    \begin{tabular}{ m{1.3cm} | m{1.5cm} | m{2.8cm} | m{1.4cm} } \hline
    \textbf{Group} & \textbf{Participant} & \textbf{Role} & \textbf{Industry Exp. (Years)}   \\ \hline

Testing & T1\label{T1} 
& 	Test Engineer	& 20
\\ \cline{2-4}

& T2\label{T2}  
& Test Engineering Manager&	41
\\ \cline{2-4}

& T3\label{T3} 
& 	Test Engineer	& 37
\\ \hline

Developer \& DevOps & D1\label{D1} 
&	Applications Developer&	22
\\ \cline{2-4}

& D2\label{D2} 
&	Middleware Engineering Manager	& 25
\\ \cline{2-4}

& D3\label{D3} 
&	Lead Software Developer	& 17
\\ \cline{2-4}

& D4\label{D4} 
&	Database Administrator	& 40
\\ \hline

IAM Analysts & I1\label{I1} 
&	IAM Business Analyst	& 19
\\ \cline{2-4}

& I2\label{I2} 
&	IAM Systems and Directory Administrator	& 26
\\ \cline{2-4}

& I3\label{I3} 
&	IAM Systems Analyst	& 37
\\ \hline

    \end{tabular}
    
    \label{table-participants}

\end{table}

\paragraph{Participant Recruitment} 
Once the DAST integration was completed, we initiated qualitative interviews to gain insights into how DAST was adopted within the Agile workflows. 
We initiated the process by sending an open invitation through the team's dedicated Slack channel, reaching out to all members to participate in the research interviews. Of the 23 team members contacted, 10 responded positively, yielding a 43\% participation rate. We then coordinated with each respondent individually to schedule virtual interviews at times that accommodated their work schedules, minimizing any impact on their ongoing projects and daily responsibilities.
As indicated in Table~\ref{table-participants}, four participants were part of the Developer \& DevOps group, three were from the Testing group, and three were from the IAM Analyst group. The distribution of participants were fairly evenly spread across the three groups in the team, allowing us to gather a well-rounded view of the DAST integration experience from the different roles.
 

\begin{table}[t]
\centering
\caption{Interview Questions}
\begin{tabular}{m{0.5cm} | m{0.5cm}  | m{6.5cm}}
\hline
\textbf{RQs} & \textbf{No.} & \textbf{Questions}  \\
\hline
\hyperref[RQ1-willing]{RQ1} & Q1\label{Q1} & ``How willing were you to include the DAST (Dynamic Application Security Testing) security scanning practice into the software development process?'' \\ 
\cline{2-3}

& Q2\label{Q2} & ``Now that it has been brought into the Agile process, how willing are you to continue DAST security scanning practice?'' \\ \hline

\hyperref[RQ2-impact]{RQ2} & Q3\label{Q3}  & ``What impact did the inclusion of this security practice have on your day-to-day work?'' \\ 
\cline{2-3}

 & Q4\label{Q4}  & ``What are the general challenges you had while incorporating security practice into your agile process?'' \\ \cline{2-3}

 & Q5\label{Q5}  & ``How did this inclusion affect the team velocity?'' \\ \cline{2-3}

 & Q6\label{Q6}  & ``Do you think the software product is more or less secure now that we have included this security practice?'' \\ \hline

\hyperref[RQ3-improvements]{RQ3} & Q7\label{Q7}  & ``What improvements can be made to better integrate security practices into Agile?'' \\ \hline

\hyperref[RQ4-compare]{RQ4} & Q8\label{Q8}  & ``How does the use of DAST compare to other security practices familiar to team members in Agile?'' \\ \hline

\end{tabular}

\label{table-interview-questions}

\end{table}

\paragraph{Interview Design} The interview questions, listed in Table~\ref{table-interview-questions}, were designed to align with our research questions exploring various dimensions of the team's experience with DAST integration in Kanban development.  Questions were open-ended to allow a broad range of responses from participants. The interview sessions were recorded using Zoom or Microsoft Teams, depending on participants' preferences. 

To answer \hyperref[RQ1-willing]{RQ1}, interview questions \hyperref[Q1]{Q1} and \hyperref[Q2]{Q2} aimed to understand the willingness of team members to initially adopt DAST and their continued interest in using it. 
By exploring the team’s openness to integrating DAST into their daily routines and Agile practices, we gained valuable data on the factors that influenced their willingness.

The next set of questions (\hyperref[Q3]{Q3}|\hyperref[Q6]{Q6}) were designed to answer \hyperref[RQ2-impact]{RQ2} by focusing on the tangible effects of DAST on the team’s workflow. \hyperref[Q3]{Q3} and  \hyperref[Q4]{Q4} explored the day-to-day challenges and disruptions caused by DAST, such as how it fits into Agile processes and whether it caused any friction with the team's existing workflow. \hyperref[Q5]{Q5} focused explicitly on the perceived impact of DAST on team velocity. \hyperref[Q6]{Q6} sought broader team members' perspectives on whether the software was more secure after implementing DAST.

Lastly, \hyperref[Q7]{Q7} and  \hyperref[Q8]{Q8} provided data to answer \hyperref[RQ3-improvements]{RQ3} and \hyperref[RQ4-compare]{RQ4} respectively, by asking the team members for their thoughts on improving the integration of security practices within Agile and how DAST compares to other security practices they were familiar with. By exploring these questions, we were able to contextualize DAST within the broader landscape of security measures and gather ideas for further improvement.

\paragraph{Participants} The ten team members who took part in our interviews brought diverse perspectives to the integration of security practices into their Agile SDLC. As shown in Table~\ref{table-participants}, the participants had significant industry experience, ranging from 17 to 41 years, with an average of 28.4 years. This wealth of experience underscores the depth of expertise they brought to the discussion on integrating security measures, such as DAST, into their Agile workflows.

\subsubsection{Data Analysis}

Interview recordings were transcribed using Zoom and Microsoft Teams functionalities to facilitate a detailed analysis. We employed the open coding process, commonly used in qualitative research\cite{fitzpatrick2013us}, to analyze the transcripts\cite{blair2015reflexive} and identify, label, and categorize data into themes or patterns without predefined codes. Two researchers conducted the analysis, each independently coding responses before merging to finalize categories, drawing insights from participant responses.

A sample of our data analysis process is depicted in Table~\ref{table-open-coding-example} for \hyperref[RQ2-impact]{RQ2}. We asked team members about their challenges while incorporating the DAST security practice into their Agile process. We highlighted parts of sentences that indicate one or more challenges. For example, \hyperref[T1]{T1} responded \textit{``One of the challenges was getting the DAST scans to work right. They are not automated in the deployment pipeline. JavaScript driver app payloads are garbage.''}. After individual analyses, we finally coded ``getting the DAST scans to work right'' as \textsl{Setup Challenges}, ``They are not automated in the deployment pipeline'' as \textsl{Lack of Automation}, etc. 
Similarly, all responses for each question were thoroughly analyzed and classified into one or more categories. We did not obtain IRB approval to share the individual responses publicly.

\begin{table*}[t]
\centering
\caption{Open coding examples for ``What are the general challenges you had while incorporating security practice into your agile process?''}
\begin{tabular}{m{1.5cm} | m{11.5cm} | m{3.4cm}}
\hline
\textbf{Participant} & \textbf{Response} & \textbf{Categories Identified} \\
\hline
\hyperref[T1]{T1} & ``One of the challenges was getting the DAST scans to work right. They are not automated in the deployment pipeline. Javascript driver app payloads are garbage.DOM faced difficulty with ZAP and other initial tools. The biggest challenge going forward would be the disconnect between the modern web and Javascript. Artificial Intelligence (AI) will get into this too, there is future
potential for an AI tool to parse web pages.'' &  Setup Challenges, Tool Limitations, Lack of Automation, Disconnect with Modern Web (JavaScript), Future Solution: AI \\
\hline

\hyperref[T2]{T2} & ''Finding the right tool that we can work with.
Our system had limitations. We could not create fake accounts for testing, and now it may be turned off.'' & Tool Compatibility, System Limitations (DUO 2FA), Testing Restrictions (No Fake Accounts) \\ \hline

\hyperref[T3]{T3} & ''Setting up a repeating card in an agile board. Earlier upper
management did not approve.'' & Cultural Challenges, Upper Management Approval\\ \hline

\hyperref[D1]{D1} & ''Finding the time bandwidth.'' & Time Constraints, Limited Bandwidth \\ \hline

\hyperref[D2]{D2} & ''I mean, there's this general challenge of parsing the report, but I don't
think that's, you know, something you're not gonna be able to avoid.'' & DAST Report Parsing Challenges, Unavoidable Challenges \\ \hline

\hyperref[D3]{D3} & ''Since the assigned engineer did all of the work, and we did not have any vulnerabilities
that needed to be fixed, there were no challenges for me. Once the assigned engineer is gone and if are to continue the DAST practice, then some of the challenges would be how to incorporate this into our pipeline, what is the right tooling, setting
the balance between which levels of alert would need to be fixed and which to ignore.'' & No Challenges, Engineer Dependence, Future Challenges: Tool Selection, Pipeline Integration, Alert Prioritization \\ \hline

\hyperref[D4]{D4} & ''I did not have any challenges since we had you working on the setup
and the configuration.'' & No Challenges, Engineer Dependence \\ \hline

\hyperref[I1]{I1} & ''I have had no general challenges while incorporating this security practice.'' & No Challenges \\ \hline

\hyperref[I2]{I2} & ''Did not encounter any challenges while incorporating this security practice
into our SDLC.'' & No Challenges \\ \hline

\hyperref[I3]{I3} & ''I haven’t had to do anything with myself, so it has not affected me except to
check them out quarterly.'' & No Challenges, Minimal Involvement \\ \hline

\end{tabular}

\label{table-open-coding-example}

\end{table*}

\subsection{Specifying Learning}

The final phase of action research involves providing general learnings based on the evaluation~\cite{petersen2014action}. Based on our findings, Section~\ref{sec:discussion} provides implications for improving DAST and general lessons learned to share insights for future teams considering DAST integration in Agile processes.

\section{Results}\label{sec:results}

This section presents the findings from our interviews. For participants, we use the T- prefix to indicate a tester, D- for developers, and I- for IAM analysts.

\subsection{\hyperref[RQ1-willing]{RQ1}: Willingness to Adopt and Continue DAST}
During our interviews, we asked participants to reflect on their willingness to adopt DAST when the team initially considered incorporating it into their project and provide insights on their willingness to continue DAST after integration.

\subsubsection{Original Willingness} 
Our interview results indicate a high willingness of Agile members to adopt DAST in their SDLC, with most interview participants expressing a positive attitude towards its integration. Of the ten interviewees, seven were very willing, two were willing, and one was unwilling to adopt DAST. \hyperref[I2]{I2} and \hyperref[D4]{D4} explicitly stated the importance of security in software development, particularly when handling sensitive information. As \hyperref[I2]{I2} noted, \emph{``Security is a critical aspect of software engineering, and incorporating DAST into our process seems like a natural step''}. \hyperref[D4]{D4} also echoed this sentiment, saying, \emph{``Security is crucial, especially when dealing with large datasets that contain sensitive information''}. 
%
 Despite the overall enthusiasm, participants raised several challenges and concerns. Practical concerns related to setup, resource allocation, and organizational support were common themes. For instance, \hyperref[T1]{T1} mentioned being \textbf{Willing} but was concerned about \textbf{Political factors} and noted that \textbf{Getting Management Approval} was a hurdle. \hyperref[T2]{T2} said that while they were \textbf{Extremely willing,} there were challenges in figuring out \emph{``how hard it is and what it will cost in money and man-hours''}. These responses highlight how concerns about implementation complexity and organizational resistance often temper willingness to adopt DAST.
Only \hyperref[D1]{D1} from the entire team was \textbf{Unwilling} to adopt DAST and 
said they were \emph{``concerned about return on investment and the development team’s time and effort that would go into it''}. These concerns suggest that while the team may enthusiastically adopt DAST, ongoing support and resources will be necessary to sustain it.

\subsubsection{Continued Willingness} All team members were highly willing to continue using DAST after integration. Seven interviewees were \textbf{Very Willing} to continue using DAST, highlighting its effectiveness in detecting vulnerabilities early and aligning well with existing workflows. For instance, \hyperref[I2]{I2} stated, \emph{``I’m very willing to continue with DAST. It fits well with the DevOps philosophy of CI/CD''}. Similarly, \hyperref[I1]{I1} emphasized the tool's practical benefits, saying, \emph{``I’m very willing to continue, especially since it has shown to be effective in identifying vulnerabilities that could compromise user identities''}. \hyperref[D4]{D4} was willing to continue after observing DAST's ability to identify specific vulnerabilities, such as SQL injections: \emph{``I’m willing to continue with it, especially since we’ve seen benefits in identifying SQL injection vulnerabilities''}.
%
No participants expressed outright unwillingness to continue using DAST. However, the four members (\hyperref[T2]{T2}, \hyperref[I2]{I2}, \hyperref[D1]{D1} and \hyperref[D3]{D3}) pointed out the need for improvements. \hyperref[D1]{D1}, for example, was \textbf{Willing} but stressed the need for more clarity in the tests and suggested involving a security specialist for better interpretation of the reports: \emph{``We need to review the set of tests that would generate alerts worth looking into. Need to be a security specialist to explain the reports''}.  
Other areas of improvement include reducing developer effort through automation and better integration into CI/CD pipelines. \hyperref[D3]{D3} said: \emph{``I am totally willing to continue, but the team has to decide. If the scans are integrated into CI/CD, it won't take much effort from the developers''}. This reflects a consensus that while team members are willing, they also seek enhancements to make DAST more efficient and less intrusive in their development workflow.

\subsection{\hyperref[RQ2-impact]{RQ2}: Perceived Impacts of DAST Integration}

\subsubsection{Impact on Day-To-Day Work}
When asked about the impact DAST integration had on participants' day-to-day work, all ten responded that it had minimal impact, with \textbf{Low/No Impact} being the most frequently observed response. This indicates that the security practice did not interfere with their core responsibilities. Participants often cited their involvement as infrequent, typically limited to reviewing quarterly reports, further emphasizing the \textbf{Occasional involvement}. For example, \hyperref[D3]{D3} noted, \emph{``It didn’t really impact my day-to-day work at all, but again, that was because [assigned engineer] did all of the heavy lifting.''} Similarly, \hyperref[D2]{D2}, \emph{``We only look at these things quarterly, almost none. 10-15 minutes every three months.''} These responses highlight participants' limited interaction with DAST due to either the low frequency of vulnerabilities found or the delegation of security tasks to other teams, such as the security engineers.
Additionally, DAST required \textbf{Negligible Effort} for most participants, with time investment restricted to occasional review of scan reports. For instance, \hyperref[I1]{I1} explained, \emph{``I spend some time reviewing the scan results to ensure there are no identity-related vulnerabilities''}, indicating that the only effort involved was a light review process. Despite the low impact and effort, some participants did report an \textbf{Increased Confidence in System Security}, mainly due to the assurance that no critical vulnerabilities were present. As \hyperref[T2]{T2} put it, \emph{``The positive impact is that we know there isn’t any big vulnerability that needs to be taken care of, increasing our level of confidence in system security''}. This balance of minimal disruption alongside a reassurance of security highlights the overall effectiveness of DAST with low overhead for day-to-day operations.

\subsubsection{Challenges}
We also asked the participants about their challenges when incorporating DAST. Six out of the ten participants mentioned that they had \textbf{No Challenge,} while others mentioned minimal challenges. This suggests integrating DAST was largely seamless for many team members, thanks to support from the engineer assigned. For instance, \hyperref[D4]{D4} noted, \emph{``I did not have any challenges since we had [assigned engineer] working on the setup and the configuration,''}. Others (\hyperref[I1]{I1}, \hyperref[I2]{I2}, \hyperref[I3]{I3}) echoed similar sentiments, indicating they faced minimal or no difficulties due to the division of tasks and reliance on specific team members to handle the setup and execution of DAST. This pattern of \textbf{Engineer Dependence} suggests that delegating security responsibilities mitigated many potential challenges, though it raises concerns about sustainability once such personnel are unavailable.

Despite the overall lack of challenges for most participants, \textbf{Time Management} and \textbf{Bandwidth Issues} emerged as hurdles for those actively involved. \hyperref[T3]{T3} and \hyperref[D1]{D1} cited finding the \textbf{Time Bandwidth} as a constraint, reflecting on the difficulty of integrating DAST alongside other development tasks in an already tight workload. Additionally, \hyperref[D2]{D2} mentioned that they encountered difficulties, such as \textbf{Report Parsing}, due to the complexity of interpreting security scan reports saying, \emph{``There’s this general challenge of sort of parsing the report, but I don’t think that’s something you’re not going to be able to avoid.''} While technical complexities were not as widespread as the time constraints or reliance on personnel, they still underscore the importance of further automating and improving DAST processes to reduce challenges for developers.

\subsubsection{Impact on Team Velocity}

We inquired about the impact of integrating DAST on team velocity. This integration had \textbf{No/Minimal Impact on Velocity}, as indicated by eight out of ten participants. The predominant sentiment reflected in the feedback is that DAST introduced some level of effort but did not alter the team's overall productivity. Three participants mentioned that relying on a dedicated engineer to handle DAST activities contributed to this perception of minimal impact. As \hyperref[D3]{D3} articulated, \emph{``There was no measurable impact of this practice on team velocity because 99.99\% of the work was done by [assigned engineer].''} This sentiment was echoed by others who reported that the separate manpower for managing DAST scans allowed the team to maintain their velocity without major disruptions. \hyperref[I2]{I2} mentioned that the low impact is due to the less frequent scans, stating, \emph{``It has a very minimal effect on the team velocity, as the scans themselves are conducted quarterly.''}

Despite the minimal impact on velocity, it is concerning that security remains an afterthought for the team, due to a \textbf{Task-Oriented} focus that prioritizes \textbf{Speed over Security}. \hyperref[I3]{I3} said the team often aims to \emph{``just sort of wanna get the job done and move the card from one column to the next,''} indicating a reluctance to pause and consider the security implications of their actions. This \textbf{Lack of Security Awareness} results in simplified measures, with many relying solely on existing tools like firewalls and failing to adopt a comprehensive security mindset. \hyperref[I3]{I3} expressed that there is a need for a cultural shift toward viewing security as an integral part of the SDLC, \emph{``If people got used to it and did it, I think it would just be part and parcel,''} reflecting the belief that security should be ingrained in daily practices rather than treated as an isolated concern. However, the challenge lies in overcoming resistance to change, as many team members may not easily adapt their mindset. \hyperref[I3]{I3} noted, \emph{``it's a big ship, and it's gonna turn slowly''}, emphasizing the \textbf{Difficulty of Instilling a Security-First Culture} amidst existing \textbf{Organizational Inertia}. Nonetheless, there is hope, as the team members are willing to advocate for security awareness and knowledge sharing, indicating a potential pathway to enhance the overall security posture.
 Overall, the impact of including DASTs on team velocity has been limited, primarily due to the structure of the Agile process and the dedicated resources allocated to security activities.
 
\subsubsection{Perceived Impact on Security}
Participants were asked if they felt the software was more secure with DAST in Agile. This inclusion had a positive impact on the product's perceived security.
Eight out of ten participants directly agreed that the product was \textbf{More Secure} after incorporating DAST. \hyperref[D3]{D3} was unsure and said that there were indirect security improvements, while \hyperref[T2]{T2} said that it is the same as before. Several members highlighted the specific benefits of DAST; these were \textbf{Increased Confidence in the Security Level}~(\hyperref[T3]{T3}, \hyperref[D3]{D3}), \textbf{Increased Security Assurance}~\hyperref[I2]{I2} and \textbf{Improved Awareness of the Current State of Security}~(\hyperref[D2]{D2}, \hyperref[T3]{T3}). Many participants noted that DAST added an \textbf{Extra Layer of Protection}~\hyperref[D4]{D4}, especially for low-risk vulnerabilities, which were sometimes overlooked in their previous processes \hyperref[I1]{I1}. For instance, \hyperref[I2]{I2} mentioned that \emph{``The DAST scans have identified vulnerabilities we might not have caught otherwise. These are critical components, and securing them has significantly reduced our risk profile and provided assurance in the security of our software systems.''} This combination of newfound confidence and reduced risk profile contributed to an overall perception of increased security among the team.

However, some participants expressed a nuanced view, pointing out the limitations of DAST in fully addressing security needs. For instance, \hyperref[D1]{D1} noted that \emph{``the security scans are missing some class of errors,''} and suggested incorporating additional tools like AI, ML, and fuzzing to cover those gaps. Despite this, most participants believed that DAST had positively impacted security, even if it didn't catch everything. \hyperref[D3]{D3} commented that \emph{``indirectly there is a security improvement anytime you introduce some software scanning process into your SDLC.''} This sentiment reflects a broader consensus that DAST has contributed to a more robust security posture, although continued improvements and complementary tools are necessary for comprehensive coverage.

\subsection{\hyperref[RQ3-improvements]{RQ3}: Improvements to DAST Integration}

We inquired about ways to improve the integration of other security practices into the Agile. Several recurring themes emerged from participant responses, particularly emphasizing the need for \textbf{Increased Testing}, \textbf{Automation}, and \textbf{Security Training}. Participants highlighted that more frequent testing and diverse types of testing, such as smoke and regression tests, are essential for identifying vulnerabilities early in the development life-cycle. \hyperref[T1]{T1} noted, \emph{``We need to move up the testing interval, i.e., make them more frequent,''} which underscores the urgency for consistent testing practices. The need for \textbf{Automation} was mentioned by \hyperref[T1]{T1}, \hyperref[D4]{D4} and \hyperref[I3]{I3}. As \hyperref[I3]{I3} articulated, \emph{``Automation to run the scans whenever new releases are made should be a part of the CI/CD pipeline.''} 
\hyperref[D3]{D3} also suggested that tools should have the capability to be integrated into the CI/CD pipeline, saying, \emph{``That would be the ultimate goal, which is to fail the pipeline if a vulnerability is found.''} \textbf{Better Feedback Loops} were deemed necessary for facilitating communication among team members, while frequent reporting was highlighted to ensure that security remains a top priority throughout the development life-cycle.
Participants emphasized that regular training sessions could improve the team's understanding of security practices within the Agile framework. \hyperref[I2]{I2} expressed, \emph{``Conducting regular security training for the DevOps team would help everyone understand the importance of security in the context of IAM.''} Furthermore, \hyperref[I1]{I1} mentioned integrating security practices with \textbf{Identity and Access Management (IAM)} checks as a crucial step in achieving a more comprehensive security posture. 

Participants also emphasized the need for \textbf{Reporting improvements}, particularly regarding consistent DAST reporting across REST APIs. \hyperref[D2]{D2} stated, \emph{``Personally, I would like to see the same sort of reporting done against our REST APIs''}. 
Additionally, \textbf{Enhancing Accessibility and Usability of Security Reports} was noted as a factor; as \hyperref[D2]{D2} expressed, \emph{``I have no idea how to look at the HTML. I have to download it and open it, which feels like an unnecessary extra step''}.
\hyperref[I3]{I3} advocated for \textbf{Anomaly Detection} and \textbf{Log Monitoring}, alongside integrating artificial intelligence (AI) for advanced analysis, saying, \emph{``We should do more sophisticated data analysis on our logs; we could use AI to process these logs''}, emphasizing the potential of AI to enhance security visibility. 
These aspects highlight desired improvements to enhance security within modern development processes.

\subsection{\hyperref[RQ4-compare]{RQ4}: Comparing DAST with Other Security Practices}

We inquired about how participants felt other familiar security practices compared to DAST. According to the responses, DAST stands out by offering \textbf{Runtime Vulnerability Detection}, a crucial advantage highlighted by four participants. Unlike \textbf{SAST}, which focuses on analyzing code for potential issues, DAST captures vulnerabilities that arise during the actual execution of the application, providing deeper insights into how an application behaves in real-world scenarios. As \hyperref[D3]{D3} noted, \emph{``DAST is more live, done at runtime, which is fantastic because the static analysis does not give us runtime vulnerabilities.''} This distinction is critical for the development team, as it complements the existing SAST practices by covering areas that static analysis alone cannot. By integrating DAST into their pipeline, teams can benefit from a \textbf{Layered Security Approach}, ensuring that security concerns are comprehensively addressed. Combining static and dynamic testing methods can strengthen overall project security by enhancing the detection of runtime and code-based vulnerabilities that might otherwise go unnoticed.

Moreover, participants also emphasized DAST's collaborative and \textbf{Proactive Nature}, which contributes to \textbf{Improving Team Communication} and the overall security workflow. For instance, \hyperref[I3]{I3} remarked, \emph{``We should track these vulnerabilities and notify the team regularly, being proactive''}, underscoring the value of \textbf{Continuous Monitoring} and real-time feedback that DAST provides. This helps teams stay aware of emerging threats and prioritize critical vulnerabilities effectively, leading to better-informed decision-making. Additionally, DAST’s ability to \textbf{Complement Other Security Practices}, such as access control and multi-factor authentication, was pointed out by \hyperref[I1]{I1} who stated, \emph{``Security scanning, like DAST, complements these practices by identifying vulnerabilities that could potentially be exploited to bypass IAM controls.''} DAST identifies application-level vulnerabilities and enhances the effectiveness of other security practices already in place, making it a crucial addition to the team's security arsenal.

\section{Discussion} \label{sec:discussion}

The findings of this study underscore the benefits and challenges of integrating DAST into Agile workflows. Through observation and interview, we explore how DAST can impact team dynamics, development velocity, security practices, and overall security posture. We found Agile team members are willing to integrate DAST into workflows and face limited challenges and impact on daily tasks with increased confidence in security. However, participants also noted improvements necessary to enhance DAST scanning and streamline DAST integration. We reflect on lessons learned to provide actionable recommendations for practitioners and broader implications for enhancing tools to streamline the integration of automated security testing in Agile development. 


\subsection{Importance of Automation and Tooling}

One of the key takeaways from this study is the critical role of automation in successfully integrating security into Agile workflows. The willingness of the team to adopt and continue using DAST highlights the increasing recognition that security cannot be sidelined in today’s fast-paced development environments. Most participants also noted DAST increased their confidence in the security of the program. 

\textit{Lesson Learned: Automate Security Scans}  Previously, the testing team of the organization adopted manual and ad hoc methods to ensure the security of the team's projects. The inefficiencies of these approaches led to the decision to seek automated security testing through DAST. Prior work also suggests security testing can be frustrating and cognitively demanding~\cite{smith2015questions}. However, automated tooling has been shown to reduce cognitive load on developers~\cite{oliveira2014s} and detect vulnerabilities efficiently and effectively~\cite{albreiki2014evaluation}, ensuring software security. Thus, Agile teams should integrate automated security tools into CI/CD pipelines to ensure continuous security scanning without manual intervention.

\textit{Lesson Learned: Assign an Engineer} Participants reported minimal challenges and impact of DAST integration to their daily work tasks, primarily due to having a dedicated engineer working on DAST integration tasks. One engineer led the efforts to explore and integrate tools, allowing other developers to continue with their normal work activities. Prior work suggests developers are unaware of security concerns and make them low priority~\cite{lopez2019hopefully}---however, Poller et al. show that security experts can enhance security in Agile organizations~\cite{poller2017can}. Agile teams considering adopting automated security testing should also consider having dedicated workers to undertake integration tasks.

\textit{Tool Recommendation: Enhanced Automation} The reliance on an assigned engineer to manage and conduct DAST integration and the expressed need for better automation point to the need to enhance DAST tooling. The more automated and seamless security practices are, the more likely they will be adopted and sustained~\cite{rajapakse2021empirical}. 
Participants noted the limitations of current DAST tools, particularly in report parsing and the interpretation of findings. These challenges indicate the need for more sophisticated tools to provide clearer, actionable insights directly within the development pipeline. Enhancements like automated report analysis, better feedback loops, and tighter integration with CI/CD project management tools could further streamline security processes and reduce efforts for developers and security specialists.

\subsection{Balancing Speed and Security in Agile}
Another recurring theme in this study was the tension between Agile and the nature of security testing. While Agile methodologies emphasize rapid delivery and iterative feedback loops, integrating comprehensive security practices can hinder speed~\cite{rindell2018aligning}. This tension was evident in the participants’ feedback, with many reporting the value of adding DAST despite the complexity of managing security scans. 

\textit{Lesson Learned: Be flexible} The team prioritized DAST integration in their CI/CD pipeline to allow security scans with limited manual intervention. However, challenges with tool limitations and manual report interpretation remain. Despite prior work suggesting automated tests should be run frequently and often~\cite{rahman2015synthesizing}, the team decided to run scans quarterly and only focus on high-priority issues. Prior work also suggests prioritizing vulnerabilities can enhance security in web applications~\cite{alptekin2020towards} and infrequent automated testing can reveal new defects while consistent replays of test cases can set a false expectation of high quality~\cite{berner2005observations}. This flexibility helped the team strike the right balance between Agile and security practices. Thus, Agile teams should be flexible in their integration of automated security testing.

\textit{Tool Recommendation: Enhance Tool Output} We observed the main challenges with adopting security scans in Agile development was parsing and understanding the generated reports. Prior work also shows understanding security reports is challenging for developers, inhibiting the security of the system~\cite{smith2020can}. Prior work has explored solutions, such as leveraging large language models (LLMs) to summarize DAST reports to make them more human-readable~\cite{arpit-dast}. Future efforts by tools and researchers can investigate additional techniques to simplify automated security reports to streamline understanding for developers in Agile teams.

\subsection{Cultural Shift Towards Security Awareness}

Our findings highlight the need for cultural shifts in how development teams view security. The team prioritized security testing and DAST integration on their Kanban board. However, we also found many team members viewed security as secondary to their primary task of delivering software quickly. In addition, challenges with ``politics'' and ``getting management approval'' inhibited willingness to adopt DAST.

\textit{Lesson Learned: Foster a Security-First Culture} Prior work suggests security is often not prioritized by software engineers~\cite{lopez2019hopefully} and highlights adversarial relationships between development and security testing teams~\cite{green2016developers}. Enhancing security in organizational culture can influence the adoption of security tools in practice~\cite{xiao2014social}. Investing in tools that provide insights from security scans can foster a security-first culture. For instance, we observed the commercial version of Burp Suite provided enhanced capabilities and support compared to the free and open-source ZAP tool. Agile is driven by feedback~\cite{kortum2019behavior}, and regular reviews of security reports, along with providing clear feedback to all stakeholders, can help ensure security remains a top priority~\cite{werlinger2009security}. By instilling a mindset that prioritizes security, organizations can encourage an environment where security is prioritized in development workflows. 

\textit{Tool Recommendation: Support Security Culture} We observed several suggestions for tooling to support a security-first culture. For instance, participants desired more understandable output and better feedback loops to facilitate communication about security issues among team members. Participants also noted the need for increased testing, automation, and training regarding DAST integration. For instance, the process of searching for DAST tools was non-intuitive, leading to restarting the process after a selected tool did not meet the team requirements due to supporting an out-of-date policy. As security becomes increasingly automated, novel features can be used to promote security in development teams. For example, this can be achieved through regular security workshops\cite{weir2023incorporating, weir2021passion}, gamified security challenges\cite{triantafyllou2022gamification}, or by integrating security checklists\cite{gilliam2003software}. 

\subsection{Comparative Strengths and Gaps of DAST}

Our study revealed DAST offers unique benefits compared to other security practices, such as the ability to identify runtime vulnerabilities that static analysis might miss, offering a comprehensive layer of protection. Participants also pointed out DAST limitations, particularly its difficulty in detecting certain classes of vulnerabilities, such as logical errors or business logic flaws, which require more contextual analysis. 

\textit{Lesson Learned: Adopt a Layered Security Approach} To address these gaps, teams may benefit from combining security testing practices. For instance, DAST can be combined with other security practices~\cite{martelleur2022security}, such as SAST~\cite{heijstek2023bridging, mateo2020combining}, 
to ensure comprehensive coverage of vulnerabilities. This layered approach--combining multiple tools and practices--addresses the the team's pre and post-deployment security concerns. Prior work suggests organizations should adopt more holistic security strategies that include static and dynamic testing~\cite{aggarwal2006integrating}, alongside continuous monitoring and security training.

\textit{Tool Recommendation: Complement Other Security Practices} In our interview results, we found participants praised DAST tools for their ability to be integrated with other security practices, including IAM analyses, multi-factor authentication, and SAST. Prior work shows the incompatibility of security tools into existing development processes is one of the main inhibitors to adoption~\cite{al2018toward}. Thus, automated security tools should aim to be compatible with additional security and development practices.

\section{Related Work}

Prior work has investigated securing web applications in modern development. For instance, researchers have proposed lightweight methods and processes to integrate security, such as incremental risk and security policy analysis~\cite{ge2006agile}, security-focused software lifecycles~\cite{williams2019secure}, continuous security testing~\cite{rangnau2020continuous}, and DevSecOps~\cite{koskinen2019devsecops}. Specifically for web application security, researchers posits the Agile Web Development with Web Framework (AWDWF)~\cite{hu2008agile} and mapping secure Scrum processes with the ISO Systems Security Engineering – Capability Maturity Model (SSE-CMM)~\cite{maier2017towards}. However, these approaches lack insights into the integration of automated security analysis tooling into modern processes.

Research suggests suggest security practices are crucial in software development to reduce vulnerabilities and prevent attack---however, integrating them into Agile-based~\cite{zaydi2024agile,riisom2018software} and CI/CD~\cite{rangnau2020continuous,bajpai2022secure} practices is hard. Prior work also highlights developer perceptions of difficulties integrating security modern development practices~\cite{arpit}.
Rindell et al.~\cite{rindell2017busting} surveyed software practitioners, outlining beliefs that traditional security processes are not compatible with iterative and incremental Agile software development. Bartsch et al.~\cite{bartsch2011practitioners} highlight challenges of integrating security practices into Agile through a literature review and interviews with practitioners from different organizations. Rahman et al.\cite{ur2016software} analyzed internet artifacts and conducted surveys to explore perceptions of integrating security in DevOps practices, and found practitioners believe DevOps activities negatively impact security.
We gather practitioner insights on challenges integrating security practices, focusing on DAST integration in Kanban and CI/CD pipelines for a development team through an action case study.


\section{Threats to Validity}

This research study on integrating DAST into modern workflows faces potential threats to validity, which may affect the interpretation of the results. \textbf{Internal validity} concerns arise from the reliance on qualitative data from interviews and observation, which may be influenced by participant bias or recall inaccuracies. Additionally, the researcher’s involvement in the team might have led to observer bias, where team members altered their behavior due to the researcher's presence. We tried to mitigate these risks by ensuring anonymous interviews and utilizing multiple coders---one external from the team---to analyze transcripts 
\textbf{External validity} is limited by the scope of the study, which focused on a single case study team using a Kanban-based Agile process. The findings may not generalize to other Agile methodologies or organizations with different team structures, cultures, or industries. While this study provides valuable insights into DAST integration in one specific context, broader generalization requires further research involving multiple teams and varied environments to validate the findings.
Lastly, \textbf{Construct and Conclusion Validity} are potential concerns due to subjective interpretations of security integration and the small sample size of interview participants. Although open coding was used to analyze qualitative data, differing perspectives on success in security integration and the potential for selection bias may affect the study's conclusions. We also only explore perceived impacts of DAST integration, and future work is needed to explore the effects of DAST effects on the security of software products.

\section{Future Work}

Future research can explore the integration of DAST across methodologies beyond Kanban. By examining how different frameworks like Scrum, Extreme Programming (XP), or hybrid Agile models incorporate security practices, researchers can identify effective strategies for various development processes---providing a broader understanding of how DAST and other security practices can be tailored to meet the needs of diverse teams and workflows. Future directions can also incorporate quantitative metrics to assess the impact of security practices on development. While this study focused on qualitative insights, combining them with quantitative data such as the number of identified vulnerabilities, the time taken to resolve them, and changes in team velocity would provide a more objective view of how security tool integration influences Agile processes. Lastly, future studies should explore the long-term sustainability of security practices in Agile environments. Longitudinal research tracking how teams continue to use and adapt to DAST over time would provide valuable insights into whether these practices remain effective and integrated into workflows after initial adoption. Expanding the scope to include multi-team and multi-industry case studies will help generalize the findings and identify best practices for different organizational contexts, industries, and security challenges.

\section{Conclusion}

We explore the integration of DAST into modern development workflows through a case study of a Kanban-based web application development team. Our findings demonstrate incorporating DAST  into the team's Kanban and CI/CD processes resulted in minimal disruption to team velocity, especially when automation and dedicated engineers are leveraged. Most team members were initially willing to adopt and continue using DAST due to its value in identifying vulnerabilities without significantly burdening their day-to-day responsibilities. The team found DAST integration aligned with iterative development in CI/CD pipelines, reduced manual effort, and improved the team’s overall confidence in system security; however, it also introduced challenges with understanding reports, prioritizing vulnerabilities, and the need for increased automation. Based on our findings, we provide actionable insights for development teams and automated security systems, emphasizing the need to balance the speed of delivery with the imperative of maintaining secure, resilient systems.


\bibliographystyle{ACM-Reference-Format}
\bibliography{sample-base}


\end{document}